\documentclass[amsmath,amssymb,twocolumn]{revtex4}
\usepackage[parfill]{parskip}    
\usepackage{graphicx}
\usepackage{amssymb}
\usepackage{epstopdf}
\usepackage{color}
\usepackage{graphicx}
\usepackage{pdfpages}

\begin{document}

\title{Minimal Geometric Deformation in asymptotically (A-)dS space-times and
the isotropic sector for a polytropic black hole}
\author{Ernesto Contreras {\footnote{On leave 
from Universidad Central de Venezuela}
\footnote{econtreras@yachaytech.edu.ec}} }
\address{Yachay Tech University, School of Physical Sciences \& Nanotechnology, 100119-Urcuqu\'i, Ecuador\\}
\author{Pedro Bargue\~no}
\address{Departamento de F\'{\i}sica, Universidad de Los Andes, Cra. 1 E No 18 A-10, Bogot\'a, Colombia}
\begin{abstract}
In the context of the Minimal Geometric Deformation method, in this paper we implement the inverse problem in a black hole 
scenario. In order to deal with an anisotropic polytropic black hole solution of the Einstein field 
equations with cosmological constant, the deformation method is slightly extended. After obtaining the isotropic sector
and the decoupler for an anisotropic (A-)dS polytropic black hole solution, we emphasize a possible relation between anisotropization/isotropization and the violation of the energy conditions.
\end{abstract}

\maketitle

\section{Introduction}\label{intro}
In recent years, the use of the Minimal Geometric Deformation (MGD) 
\cite{randall1999a,randall1999b,antoniadis1990,antoniadis1998,ovalle2008,ovalle2009,ovalle2010,casadio2012,ovalle2013,ovalle2013a,
casadio2014,casadio2015,ovalle2015,casadio2015b,
ovalle2016, cavalcanti2016,casadio2016a,ovalle2017,
rocha2017a,rocha2017b,casadio2017a,ovalle2018,estrada2018,ovalle2018a,lasheras2018,gabbanelli2018,
sharif2018,fernandez2018,fernandez2018b,contreras2018,estrada,contreras2018a,morales}
as a systematic and powerful
method to obtain new and relevant solutions of the Einstein field equations, has considerable increased \cite{ovalle2017,ovalle2018,estrada2018,ovalle2018a,lasheras2018,gabbanelli2018,
sharif2018,contreras2018,contreras2018a,rincon2018,ovalleplb}. 
For example, the method has allowed to induce local anisotropies in spherically
symmetric systems leading to both more realistic interior solutions of compact objects 
\cite{lasheras2018,gabbanelli2018} and hairy black holes \cite{ovalle2018a}. More recently, the 
method has been
extended to solve the inverse problem \cite{contreras2018a}, namely, given any anisotropic solution of the Einstein field
equations it is possible to recover the isotropic source and the decoupler matter content which,
after gravitational interaction, led to the anisotropic configuration. In that work, it was found
that, for a simple anisotropic solution violating all the energy conditions, the free parameters involved in the MGD
can be fitted in such a manner that both the isotropic source and the matter decoupler 
content satisfy all the energy conditions. The importance of this result lies in the fact
that the inverse problem allowed to interpret the MGD as some kind of mechanism which leads to the apparition of exotic 
matter after gravitational interaction of well behaved matter content.

In the same spirit of Ref \cite{contreras2018a}, it could be interesting to explore if the same duality exotic/non--exotic matter content 
occurs in other scenarios after the application of the inverse MGD problem. 
To be more precise, we could study if such a duality persists in situations where the starting point is a solution sourced by a matter content satisfying all the energy conditions.
In order to do so, in 
this work we implement the inverse problem program in a polytropic black hole (BH)
originally studied in reference \cite{setare} and extended to the scale--dependent scenario in reference \cite{contreras2018b}. 
As it will be shown in the rest of the manuscript, the choice of such a system is twofold: first, to extend the MGD in order to 
deal with Einstein field equations with cosmological constant and, second, to
implement the inverse problem in an anisotropic system which satisfies all the energy conditions.

This work is organized as follows. In the next section we briefly review the MGD-decoupling method.
In section \ref{isotropation} we develop the method to obtain the generator of any anisotropic solution 
of the Einstein Field Equations and then we implement the method for a polytropic BH solution 
in section \ref{isoBH}. The last section is devoted to final comments and conclusion.

\section{Einstein Equations with cosmological constant and extended MGD--decoupling}\label{mgd}
With the purpose to extend the MGD in order to consider the Einstein field equations with cosmological constant,
we write
\begin{eqnarray}\label{einsorig}
R_{\mu\nu}-\frac{1}{2}R g_{\mu\nu}+\Lambda g_{\mu\nu}=-\kappa^{2}T_{\mu\nu}^{tot},
\end{eqnarray}
and we assume that the total energy-momentum tensor is given by
\begin{eqnarray}\label{total}
T_{\mu\nu}^{(tot)}=T_{\mu\nu}^{(m)}+\alpha\theta_{\mu\nu},
\end{eqnarray}
where $\alpha$ is a constant. As usual, the energy--momentum tensor for a perfect fluid 
$T^{\mu(m)}_{\nu}=diag(-\rho,p,p,p)$ and the decoupler matter content $\theta_{\mu\nu}$ 
interact only gravitationally, 
\begin{eqnarray}
\nabla_{\mu}T^{\mu(m)}_{\nu}=\nabla_{\mu}\theta^{\mu}_{\nu}=0.
\end{eqnarray}
In what follows, we shall work with spherically symmetric space--times with a line element
parametrized as
\begin{eqnarray}\label{le}
ds^{2}=e^{\nu}dt^{2}-e^{\lambda}dr^{2}-r^{2}d\Omega^{2},
\end{eqnarray}
where $\nu$ and $\lambda$ are functions of the radial coordinate $r$ only. 
Considering Eq. (\ref{le}) as a solution of the Einstein Field Equations, we obtain
\begin{eqnarray}
\kappa ^2 \tilde{\rho}&=&\frac{e^{-\lambda} \left(r \lambda '-1\right)}{r^2}
+\frac{\Lambda  r^2+1}{r^2}\label{eins1}\\
\kappa ^2 \tilde{p}_{r}&=&\frac{e^{-\lambda } \left(r \nu '+1\right)}{r^2}-\frac{1}{r^2}
-\Lambda\label{eins2}\\
\kappa ^2 \tilde{p}_{\perp}&=&-\Lambda -\frac{e^{-\lambda} \left(\left(r \nu '+2\right)\left(
\lambda '-\nu '\right)-2 r \nu ''\right)}{4 r}\label{eins3}
\end{eqnarray}
where the prime denotes derivation with respect to the radial coordinate and we have defined
\begin{eqnarray}
\tilde{\rho}&=&\rho+\alpha\theta^{0}_{0}\label{rot}\\
\tilde{p}_{r}&=&p-\alpha\theta^{1}_{1}\label{prt}\\
\tilde{p}_{\perp}&=&p-\alpha\theta^{2}_{2}.\label{ppt}
\end{eqnarray} 

The next step consists in decoupling the Einstein Field
Equations (\ref{eins1}), (\ref{eins2}) and (\ref{eins3}) by performing
\begin{eqnarray}\label{def}
e^{-\lambda}=\mu +\alpha f,
\end{eqnarray}
where $f$ is the geometric deformation undergone by the radial metric
component $\mu$, ``controlled'' by the free parameter $\alpha$. By doing so, we obtain
two sets of differential equations: one describing an isotropic system sourced by
the conserved energy--momentum tensor of a perfect fluid $T^{\mu(m)}_{\nu}$ an the other
set corresponding to quasi--Einstein Field Equations sourced by $\theta_{\mu\nu}$. 
After taking into account that the cosmological constant can be interpreted as some kind of isotropic fluid,
we include the $\Lambda$--term in the isotropic sector and
we obtain
\begin{eqnarray}
\kappa^{2}\rho &=&\frac{\Lambda  r^2-r \mu '-\mu +1}{r^2}\label{iso1}\\
\kappa ^2 p&=&\frac{-\Lambda  r^2+r \mu (r) \nu '+\mu -1}{r^2}\label{iso2}\\
\kappa ^2 p&=&\frac{2 \mu '+2 r \mu \nu ''+r \mu  \nu '^2+2 \mu \nu '}{4  
r}+\nonumber\\
&&+\frac{\mu'\nu'}{4}+\Lambda
,\label{iso3}
\end{eqnarray}
for the perfect fluid and
\begin{eqnarray}
\kappa ^2  \theta^{0}_{0}&=&-\frac{r f'+f}{r^{2}}\label{aniso1}\\
\kappa ^2 \theta^{1}_{1}&=&-\frac{r f \nu '+f}{r^{2}}\label{aniso2}\\
\kappa ^2\theta^{2}_{2}&=&-\frac{f' \left(r \nu '+2\right)+f \left(2 r \nu ''+r \nu '^2+2 \nu '\right)}{4 r},\label{aniso3}
\end{eqnarray}
for the anisotropic system \footnote{In what follows we shall assume $\kappa^{2}=8\pi$ }. We would like to emphasize that that the addition of the cosmological constant only affects 
the isotropic sector because  Eqs. (\ref{aniso1}), (\ref{aniso2})
and (\ref{aniso3}) remain unchanged. At this point, we are ready to implement the inverse problem program.

\section{MGD-decoupling: the inverse problem}\label{isotropation}
In a previous work \cite{contreras2018a}, the inverse MGD problem was solved after realizing that,
given any anisotropic solution
with metric functions $\{\nu,\lambda\}$, matter content $\{\tilde{\rho},\tilde{p}_{r},\tilde{p}_{\perp}\}$ and definitions
given by Eqs. (\ref{rot}), (\ref{prt}) and (\ref{ppt}), the following constraint must be
satisfied:
\begin{eqnarray}\label{constraint}
\tilde{p}_{\perp}-\tilde{p}_{r}=-\alpha(\theta^{2}_{2}-\theta^{1}_{1}).
\end{eqnarray}
It should be noted that the above constraint allows us to obtain a differential equation
for the decoupling function $f$ in terms of well known quantities of anisotropic solution. To be more precise, with this constraint we do not need any artificial equation of state for the $\theta$'s components.
It is worth mentioning that, in the case of Einstein equations with cosmological constant,
the previous constraint leads to the same result obtained in Ref. \cite{contreras2018a} because there is no contribution 
of $\Lambda$. More precisely, after subtracting  Eqs. (\ref{prt}) and (\ref{ppt}),
the cosmological constant disappears and the solution remains the same. 
In this sense, the combination of Eqs. (\ref{aniso2}) 
and (\ref{aniso3}) with the constraint (\ref{constraint}) leads to a differential equation 
for the decoupling function $f$ given by
\begin{eqnarray}\label{f}
f'-\mathcal{F}_{1}f=\mathcal{F}_{2},
\end{eqnarray}
where we have introduced the auxiliary functions $\mathcal{F}_{1}$ and $\mathcal{F}_{2}$  as
\begin{eqnarray}
\mathcal{F}_{1}&=&\frac{4-r \left(2 r \nu ''+\nu ' \left(r \nu '-2\right)\right)}{r \left(r 
\nu '+2\right)}\\
\mathcal{F}_{2}&=&\frac{e^{-\lambda } \left(r 
\left(-\lambda ' \left(r \nu '+2\right)+2 r \nu ''+\nu ' \left(r \nu '-2\right)\right)\right.}{\alpha  r \left(r \nu '+2\right)}\nonumber\\
&& +\frac{\left.4 e^{-\lambda }\left(e^{\lambda}-1\right)\right)}{\alpha  r \left(r \nu '+2\right)}.
\end{eqnarray}
From Eq. (\ref{f}), it is straightforward to derive that the deformation function, $f$, is given by
\begin{eqnarray}\label{fs}
f(r)=e^{\int^r \mathcal{F}_{1} \, du} \int^r \mathcal{F}_{2} e^{-\int^{w}\mathcal{F}_{1} \, du} \, dw.
\end{eqnarray}
The next step consists in obtaining the metric function, $\mu$,
by replacing Eq. (\ref{fs}) in the geometric
deformation relation (\ref{def}). We obtain 
\begin{eqnarray}\label{mu}
\mu=e^{-\lambda}-\alpha  e^{\int^r \mathcal{F}_{1} \, du} \int^r
\mathcal{F}_{2} e^{-\int^{w} \mathcal{F}_{2} \, du} \, dw.
\end{eqnarray}

Now, from  Eqs. (\ref{iso1}), (\ref{iso2}) and (\ref{iso3}), the matter content for the 
isotropic system reads
\begin{align}
\rho&=\Lambda+\mathcal{G}_{1}+\alpha  \mathcal{G}_{2} e^{\int^r \mathcal{F}_{1} \, du} 
\int^r \mathcal{F}_{2} e^{-\int^{w} \mathcal{F}_{1} \, du} \, dw\label{rho}\\
p&=-\Lambda+\mathcal{G}_{3}-\alpha \mathcal{G}_{4} e^{\int^r \mathcal{F}_{1}du} 
\int^r \mathcal{F}_{2}e^{-\int^{w}\mathcal{F}_{1}du}dw\label{pr}.
\end{align}
where we have introduced four additional auxiliary functions as
\begin{eqnarray}
\mathcal{G}_{1}&=&\frac{r e^{-\lambda} \left(\left(e^{\lambda}-3\right) \nu '+2 r \nu ''
+r \nu '^2\right)}{8 \pi  r^2 \left(r \nu '+2\right)}\nonumber\\
&&+\frac{6 e^{-\lambda } \left(e^{\lambda }-1\right)}{8 \pi  r^2 \left(r \nu '+2\right)}
\\
\mathcal{G}_{2}&=&\frac{6-r \left(2 r \nu ''+\nu ' \left(r \nu '-3\right)\right)}{8 \pi  r^2 \left(r \nu '+2\right)}\\
\mathcal{G}_{3}&=&\frac{e^{-\lambda } \left(-e^{\lambda }+r \nu '+1\right)}{8 \pi  r^2}\\
\mathcal{G}_{4}&=&\frac{r \nu '+1}{8 \pi  r^2}.
\end{eqnarray}

To determine the decoupler matter content we simply replace Eqs (\ref{fs}) and (\ref{mu}) in (\ref{aniso1}), 
(\ref{aniso2}) and (\ref{aniso3}) to obtain
\begin{eqnarray}
\theta^{0}_{0}&=&
-\frac{(r\mathcal{F}_{1}+1) e^{\int^r \mathcal{F}_{1} \, du} \int^r \mathcal{F}_{2} e^{-\int^w \mathcal{F}_{1} \, du} \, dw}{r^2}\nonumber\\
&&+\frac{\mathcal{F}_{2}}{r}\label{teta00}\\
\theta^{1}_{1}&=&
-\frac{\mathcal{H}_{1} e^{\int^r \mathcal{F}_{1} \, du} \left(\int^r \mathcal{F}_{2}
 e^{-\int^w \mathcal{F}_{1} \, du} \, dw\right)}{r^2}\label{teta11}\\
\theta^{2}_{2}&=&
-\frac{ e^{\int^r \mathcal{F}_{1} \, du} \left(\int^r \mathcal{F}_{2} 
e^{-\int^w \mathcal{F}_{1} \, du} \, dw\right)}{4 r}\label{teta22}
\nonumber\\
&&+\frac{\mathcal{F}_{2} 
\left(\mathcal{H}_{1}+1\right)}{4 r},
\end{eqnarray}
where
\begin{eqnarray}
\mathcal{H}_{1}&=&1+r \nu '\\
\mathcal{H}_{2}&=&\left(\left(r \nu '+2\right) \left(\mathcal{F}_{1}+\nu '(r)\right)+2 r \nu ''(r)\right).
\end{eqnarray}

At this point some comments are in order. First, note that the cosmological constant affects only
the isotropic sector, as previouly said, whereas the anisotropic one remains unchanged with respect to the reported
in Ref. \cite{contreras2018a}. Second, as reported in Ref. \cite{contreras2018a}, Eqs. (\ref{mu}), (\ref{rho}) and (\ref{pr}) determine the isotropic 
generator $\{\mu,\rho,p\}$ 
and Eqs. (\ref{teta00}), (\ref{teta11}) and (\ref{teta22})
determine the decoupler matter content $\{\theta^{0}_{0},\theta^{1}_{1},\theta^{2}_{2}\}$ once any anisotropic solution $\{\nu,\lambda,\tilde{\rho},\tilde{p}_{r},\tilde{p}_{\perp}\}$ is provided.

In the next section, we shall briefly review the mains aspects of the polytropic BH
reported in reference \cite{setare} and then we shall implement the method
to obtain its isotropic generator and decoupler matter content.

\section{Isotropic sector of a polytropic BH solution}\label{isoBH}
Let us start this section sumarizing the main results obtained in Ref. \cite{setare}. 
The line element is parametrized as
\begin{align}\label{metric}
&ds^{2}=\left(\frac{r^2}{L^2}-\frac{2 M}{r}\right)dt^{2}-\left(\frac{r^2}{L^2}-\frac{2 M}{r}
\right)^{-1} d r^{2}-r^{2}d \Omega^{2},
\end{align}
with $L^{2}=-3/\Lambda$. Imposing this metric as a solution of the Einstein
field equations with cosmological constant (see Eq. (\ref{einsorig})) with 
$T^{\mu\nu}=diag(-\tilde{\rho},\tilde{p}_{r},\tilde{p}_{\perp},\tilde{p}_{\perp})$ it 
is obtained that
\begin{align}
&\tilde{\rho}=-\tilde{p}_{r}=\frac{1}{8\pi r^{2}}, \label{rhop}
\\
&\tilde{p}_{\perp}=\tilde{p}_{\perp}=0.\label{pdos}
\end{align}
It is remarkable that $T_{\mu\nu}$ fulfils
all the energy conditions
\begin{align}\label{fullmatter}
&\tilde{\rho}\ge0,          
\hspace*{2.79cm}
\tilde{\rho}+\tilde{p}_{i}\ge0,\\
&\tilde{\rho}+\sum\limits_{i}\tilde{p}_{i}\ge 0,
\hspace*{1.5cm}
\tilde{\rho}+\tilde{p}_{i}\ge0,\\
&\tilde{\rho}\ge|\tilde{p}_{i}|,
\end{align}
which are referred as the weak, strong and dominant energy conditions, respectively. It is worth mentioning that the solution is singular at $r=0$ and has both a Killing and a causal horizon located at $r=2^{1/3}M^{1/3}L^{2/3}$.

In what follows, we shall implement the inverse problem protocol to obtain the isotropic generator
and the decoupler matter content associated to this polytropic BH solution. From Eq. (\ref{fs}), the 
decoupling function, $f$, reads
\begin{equation}\label{fr}
f(r)=e^{\nu} \left(\frac{\left(2^{4/3} L^{2/3}\right) \left(2 \sqrt{3} \tan ^{-1}A+\log B\right)}{3 \alpha  M^{2/3}}+6 c_1\right)
\end{equation}
where
\begin{eqnarray}
A&:=&\frac{2 \sqrt[3]{2} r}{\sqrt{3} \sqrt[3]{L^2 M}}+\frac{1}{\sqrt{3}}\\
B&:=&\frac{\left(\frac{2 M r^3}{L^2}\right)^{1/3}+\left(\frac{2 r^3}{L^2}\right)^{2/3}+M^{2/3}}{3 \left(\sqrt[3]{M}-\sqrt[3]{2} \sqrt[3]{\frac{r^3}{L^2}}\right)^2}\\
e^{\nu}&=&\frac{2 M}{r}-\frac{r^2}{L^2}
\end{eqnarray}

Replacing Eq. (\ref{fr}) in the decoupling equation, $e^{-\lambda}=\mu+\alpha f$, we obtain
an expression for the metric function $\mu$ : 
\begin{eqnarray}\label{mufinal}
\mu &=& -\frac{\left(2 L^2 M-r^3\right) \left(2\ 2^{2/3} \sqrt{3} \tan ^{-1}(A)+2^{2/3} \log (B)\right)}{6 r \left(L^2 M\right)^{2/3}}\nonumber\\
&&-\frac{3 \left(3 \alpha  c_1-2\right) \left(2 L^2 M-r^3\right)}{6 L^2 r}.
\end{eqnarray}
Using the above result in Eqs. (\ref{rho}) and (\ref{pr}), the isotropic fluid reads
\begin{align}
\rho&=\frac{\frac{12 r^3}{L^2}-15 M}{8 \pi  r^2 \left(\frac{6 r^3}{L^2}-3 M\right)}
+9 \alpha  c_1 C \left(\frac{M}{L}\right)^{2/3}\nonumber\\
&+C \left(2\ 2^{2/3} \sqrt{3} \tan ^{-1}(A)+2^{2/3} \log (B)\right)\\
p&=\frac{2 \sqrt{3} \tan ^{-1}(A)+\log (B)}{\sqrt[3]{2} 8 \pi  \left(L^2 M\right)^{2/3}}+\frac{9 \alpha  c_1 r^2}{16 \pi  L^2 r^2}-\frac{1}{8 \pi  r^2}
\end{align}
where
\begin{equation}
C:=\frac{9 \left(M-\frac{2 r^3}{L^2}\right)}{48 \pi  (L M)^{2/3} \left(\frac{6 r^3}{L^2}-3 M\right)}
\end{equation}

At this point some comments are in order. First, the horizon of the solution is located at
\begin{eqnarray}
r_H=\sqrt[3]{2} L^{2/3} \sqrt[3]{M},
\end{eqnarray}
which coincides with that of the anisotropic solution of Eq. (\ref{metric}). We think this is remarkable:
starting from an isotropic solution with an event horizon, one can form, following the MGD, the anisotropic extension of
the BH but mantaining the location of the horizon. In this sense, the entropy of both anisotropic and isotropic BHs are indistinguishable .
Second, a critical radius appears at
\begin{eqnarray}
r_c=\frac{L^{2/3} \sqrt[3]{M}}{\sqrt[3]{2}},
\end{eqnarray}
where the solution diverges. Even more, an analysis of the curvature scalar, which is given by
\begin{eqnarray}
R&=&-12 \alpha  c_1-\frac{6 r}{L^2 M-2 r^3}+\frac{12}{L^2}-\frac{8}{r^2}\\
&&+2\frac{2^{2/3} \left(2 \sqrt{3} \tan ^{-1}(A)+\log (B)\right)}{L^{4/3} M^{2/3}},
\end{eqnarray}
reveals that this critical radius is a real singularity. It is worth noticing that, the location of 
$r_{c}$ is independent of the free parameters of the theory unlike the obtained in 
\cite{ovalle2018a} and, in this sense, we can not control its location. However, the critical radius is less than the horizon radius which implies that the singularity is hidden inside the BH horizon. Third, the energy density reach a maximum at $r_M=\sqrt[3]{3 \sqrt{\frac{5}{2}} L^2 M+5 L^2 M}$ and 
its asymptotic behaviour is given by
\begin{eqnarray*}
&&\lim\limits_{r\to r_{c}}\rho\to -\infty\\
&&\lim\limits_{r\to\infty}\rho=\frac{1}{16} \left(\frac{6 c_{1}\alpha}{\pi }-\frac{2^{2/3} \sqrt{3}}{L^{4/3} M^{2/3}}-\frac{6}{\pi  L^2}\right)
\end{eqnarray*}
and at the horizon
\begin{eqnarray*}
\rho(r_{H})=\frac{-9.49286 L^{2/3}+6 c_{1}\alpha L^2 M^{2/3}-6 M^{2/3}}{16 \pi  L^2 M^{2/3}}.
\end{eqnarray*}
At this point a couple of comments are in order.  First, note that near the critical radius, the apparition of exotic matter is unavoidable. Second, as far as $r\to\infty$
the choice of the free parameters leads to a negative, positive or even vanishing energy density.
Even more, given that $\rho$ is a continuous function that reaches a maximum at $r_{M}$, the
energy density profile could be endowed with one, two or with no real roots at all. It is worth noticing that when no 
real roots appear, the exotic behaviour is present in all the space--time. When two real roots are allowed, two exotic sectors 
appear, one near the critical point and the other for some $r>r_{c}$. Finally, for only one real root, the apparition
of the exotic matter can be minimized. Specifically, one condition for the apparition of only one
real root leads to
\begin{eqnarray}
c_1\alpha= \frac{6 M^{2/3}+9.49286 L^{2/3}}{6 L^2 M^{2/3}},
\end{eqnarray}
which implies that the root is located at the horizon, {\it i. e.},
\begin{eqnarray}
\rho(r_{H})=0
\end{eqnarray}
For this value fo $c_{1}\alpha$, the energy density reach a positive value at infinity given by
\begin{eqnarray*}
\rho\approx\frac{0.0170132}{L^{4/3} M^{2/3}}
\end{eqnarray*}

\begin{figure}[h!]
\includegraphics[scale=0.5]{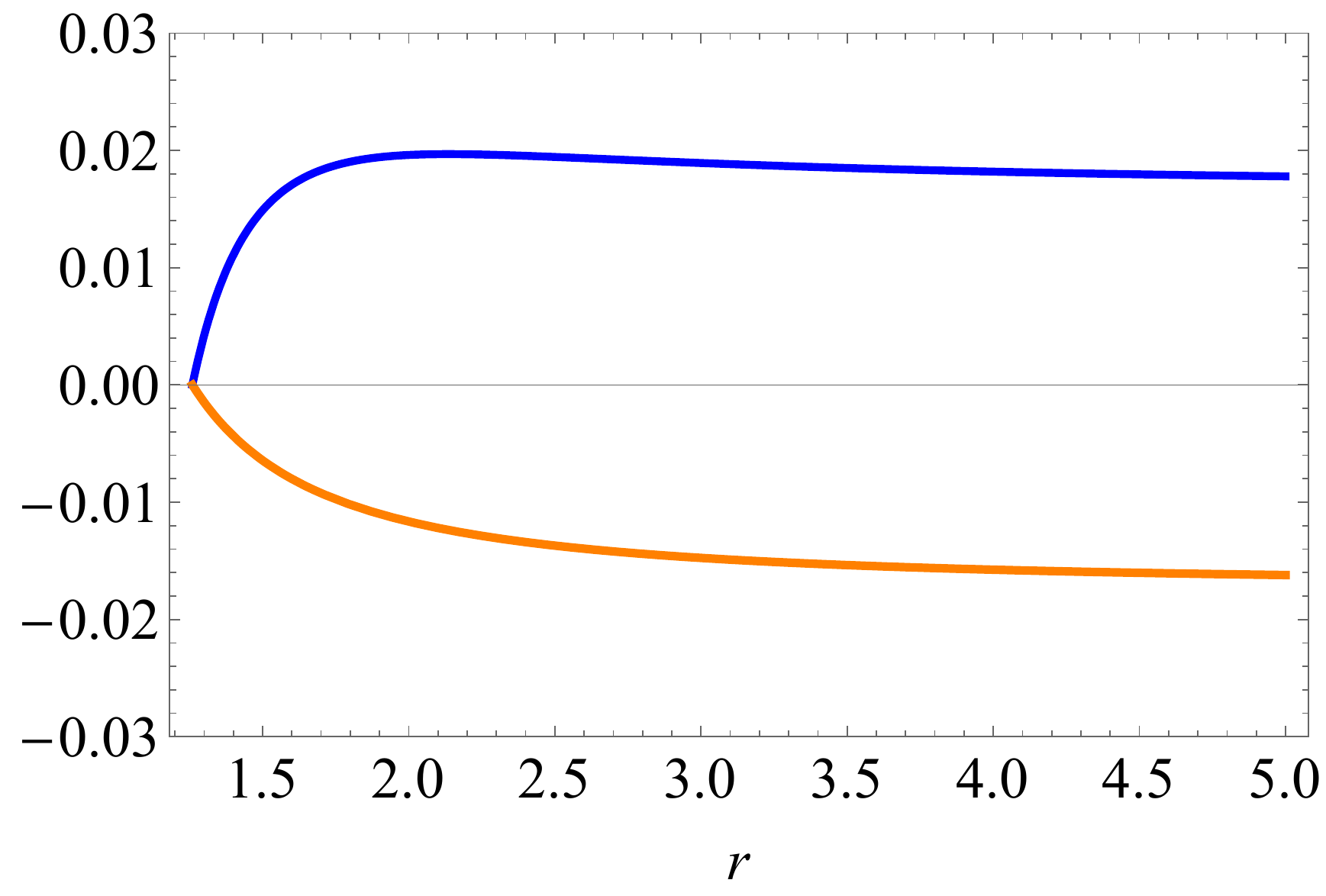}
\caption{Energy density (blue line) and pressure (orange line) profile for $L=1$}
\label{fig}
\end{figure}

In Fig. (\ref{fig}) we show the behaviour of the energy density (blue line) and the pressure (orange line)
for $r>r_{H}$, $c_1\alpha=\frac{6 M^{2/3}+9.49286 L^{2/3}}{6 L^2 M^{2/3}}$
and $M=L=1$, revealing that, for $r>r_{H}$, all the energy conditions are satisfied.

The results obtained so far could be interpreted as follows. The isotropic sector of the 
polytropic BH reported in \cite{setare} corresponds to a BH 
solution hiding exotic matter in its interior. What is more, the new isotropic BH solution obtained
in this work could be thought as the remnant of a gravitational collapse involving exotic matter which is the responsible of sustain traversable wormholes \cite{Morris1987,Morris1988,Visserbook,Lobo2017}.

\section{Conclusions}\label{remarks}
In this work we have extended the Minimal Geometric Deformation approach when a cosmological constant is present
showing that, in this case, only the isotropic sector is modified.
In particular, the inverse problem in the context of polytropic black holes has been explored, obtaining the isotropic
sector from which an anisotropic (A)--dS polytropic black hole is obtained. Moreover, the isotropic sector contains
a singularity which does not depends on any of the free parameters of the deformation approach. This singularity is hidden inside
an event horizon, which remarkably coincides with that of the anisotropic black hole. Finally, we have noted that the isotropic
sector is deeply linked with the appearance of exotic matter, although it can be located inside the horizon. In this sense,
this work shows a nice example of how one could, in principle, control the energy conditions by tuning the isotropy/anisotropy
of a black hole solution.

\end{document}